\documentclass[reprint,preprintnumbers,superscriptaddress,amsmath,amssymb,prb]{revtex4-1}
\usepackage{braket}
\usepackage{miller}
\usepackage{changes}
\usepackage[textsize=tiny]{todonotes}
\usepackage{graphicx}
\usepackage[utf8]{inputenc}
\usepackage{multirow}
\usepackage{float}
\usepackage{bm}
\usepackage{verbatim}
\usepackage{xfrac}
\usepackage{array}
\newcolumntype{C}[1]{>{\centering\let\newline\\\arraybackslash\hspace{0pt}}m{#1}}
\newcolumntype{L}[1]{>{\raggedright\let\newline\\\arraybackslash\hspace{0pt}}m{#1}}
\usepackage{siunitx}

\begin{document}
	
	\title{High-fidelity and ultrafast initialization of a hole-spin bound to a Te isoelectronic center in ZnSe}
	
	\author{P. St-Jean}
	\email{philippe.st-jean@polymtl.ca}
	\affiliation{Polytechnique Montréal, Montréal, QC, H3C 3A7, Canada}
	
	\author{G. Éthier-Majcher}
	\affiliation{Polytechnique Montréal, Montréal, QC, H3C 3A7, Canada}
	
	\author{R. André}
	\affiliation{Nanophysics and Semiconductor Group, Institut Néel, CEA/CNRS/Université Joseph Fourier, 25 rue des Martyrs, 38042 Grenoble, France}
	
	\author{S. Francoeur}
	\affiliation{Polytechnique Montréal, Montréal, QC, H3C 3A7, Canada}
	
	\date{\today}
	\begin{abstract}
		We demonstrate the optical initialization of a hole-spin qubit bound to an isoelectronic center (IC) formed by a pair of Te impurities in ZnSe, an impurity/host system providing high optical homogeneity, large electric dipole moments, and potentially advantageous coherence times. The initialization scheme is based on the spin-preserving tunneling of a resonantly excited donor-bound exciton to a positively charged Te IC, thus forming a positive trion. The radiative decay of the trion within less than \SI{50}{\pico\second} leaves a heavy-hole in a well-defined polarization-controlled spin state. The initialization fidelity exceeds $98.5~\%$ for an initialization time of less than \SI{150}{\pico\second}.

	\end{abstract}	
	\maketitle
	
	
	Interfacing long-lived solid-state qubits with optical fields is the cornerstone of long-distance transmission of information inside quantum networks\cite{Kimble2008}. Optically addressable hole-spins bound to semiconductor nanostructures are promising candidates for building such quantum interfaces\cite{Awschalom2013, Morton2011, Gerardot2008, Godden2012, DeGreve2011}. Indeed, the energy splitting between hole and trion states typically resides within the optical or near-infrared region of the electromagnetic spectrum, allowing for an efficient mapping of their quantum states onto photon polarization states that can be transmitted through optical networks. In addition, the coherence time of hole-spins is usually an order of magnitude longer than that of electrons, as p-type wave-functions mitigate the hyperfine interaction with nuclear spins \cite{Eble2009, Fallahi2010}.
	
	However, the scalability of quantum networks using optically addressable spins is compromised by the challenging task of finding an emitter that provides both a strong electric dipole moment and a high optical homogeneity. The first allows rapid and high-fidelity optical initialization, control, and single-shot read-out of quantum states. The second facilitates scalabitily by alleviating implementation complexity and providing easier means to interface qubits together, to a cavity, and to an external driving field \cite{Imamoglu1999}. On the one hand, systems exhibiting strong electric dipole moments, such as charges confined to epitaxial quantum dots, usually suffer from important inhomogeneous broadening. On the other hand however, highly homogeneous systems, such as impurity-bound charges (e.g. defects in diamond and SiC), usually exhibit electric dipole moments at least one order of magnitude weaker than semiconductor nanostructures\cite{Faraon2012, Hain2014}. It is in this context that impurity-bound excitonic complexes in various semiconductor hosts are actively investigated \cite{Sleiter2013,Dotti2015}.
	
	Lesser known in the context of quantum information, isoelectronic centers (ICs) in semiconductors provide both key advantages. Formed from one or few isoelectronic impurities, ICs provide the exceptionally high optical homogeneity of atomic-size systems with an inhomogeneous broadening determined by the quality of the host crystal. ICs bind single electrons or holes, and multiple excitonic complexes like excitons, trions, and biexcitons \cite{Marcet2010a} with electric dipole moments as strong as in quantum dots\cite{Ethier-Majcher2014}. Although several IC systems have been studied over the past decades,\cite{Francoeur2004, Thomas1966, Ikezawa2007, Jo2013, Muller2006, Weber1982} pairs of Te atoms (dyads) in ZnSe offer unique advantages for implementing optically addressable spin qubits. Most isotopes of Zn, Se, and Te have vanishing nuclear spins, thereby favoring long spin relaxation and coherence times in conditions where hyperfine interaction is the dominant decoherence mechanism. Te dyads primarily bind holes, which further reduces hyperfine interaction with nuclear spins, and positive trions that can be used, through their optical selection rules\cite{Marcet2010}, as intermediate states to initialize a hole-spin in a well-defined state. Contrary to single-atom ICs and dopants, lower dyad concentrations are readily achieved eliminating the need for sub-micron patterning, which facilitate their integration in nanophotonic devices such as optical cavities and waveguides.
	
	In this letter, we demonstrate fast on-demand optical initialization of a hole-spin bound to an IC formed by a Te dyad inside a ZnSe host. In doing so, we also demonstrate a novel initialization scheme, proper to IC systems, based on the efficient tunneling of an exciton from a resonantly excited donor-bound state to a single hole bound to an IC. The rapid radiative decay of the trion leaves a single hole bound to the dyad in a well-defined spin state, with a fidelity given by the degree of polarization of the emission. Under favorable excitation conditions, this fidelity exceeds $\mathrm{98.5~\%}$.
	
	
	The samples investigated were grown by molecular beam epitaxy on a GaAs-(100) substrate and consists of a single Te-doped plane at the center of a \SI{80}{\nano\meter}-thick ZnSe layer. The estimated density of Te atoms is $\mathrm{2500~\mu m^{-2}}$ which leads, assuming a random distribution of non-interacting Te \cite{Thomas1966} to a dyad density of $\mathrm{4 ~\mu m^{-2}}$. Micro-photoluminescence measurements were performed in a $\mathrm{1~\mu m^{2}}$-resolution confocal microscope at $T=4~K$. Excitation was provided by a frequency-doubled tunable 1-ps Ti-sapphire laser and emission was analyzed with a spectrometer coupled to an avalanche photodiode providing a spectral resolution of 60~$\mu$eV and a temporal resolution of 80~ps. Time-integrated (CW) measurements were performed with a 405~nm laser diode and a CCD camera. The polarization of the excitation and the detection were both controlled with a $\lambda/4$ wave-plate and a polarizer.
	
\begin{figure}
		\includegraphics[width=83mm]{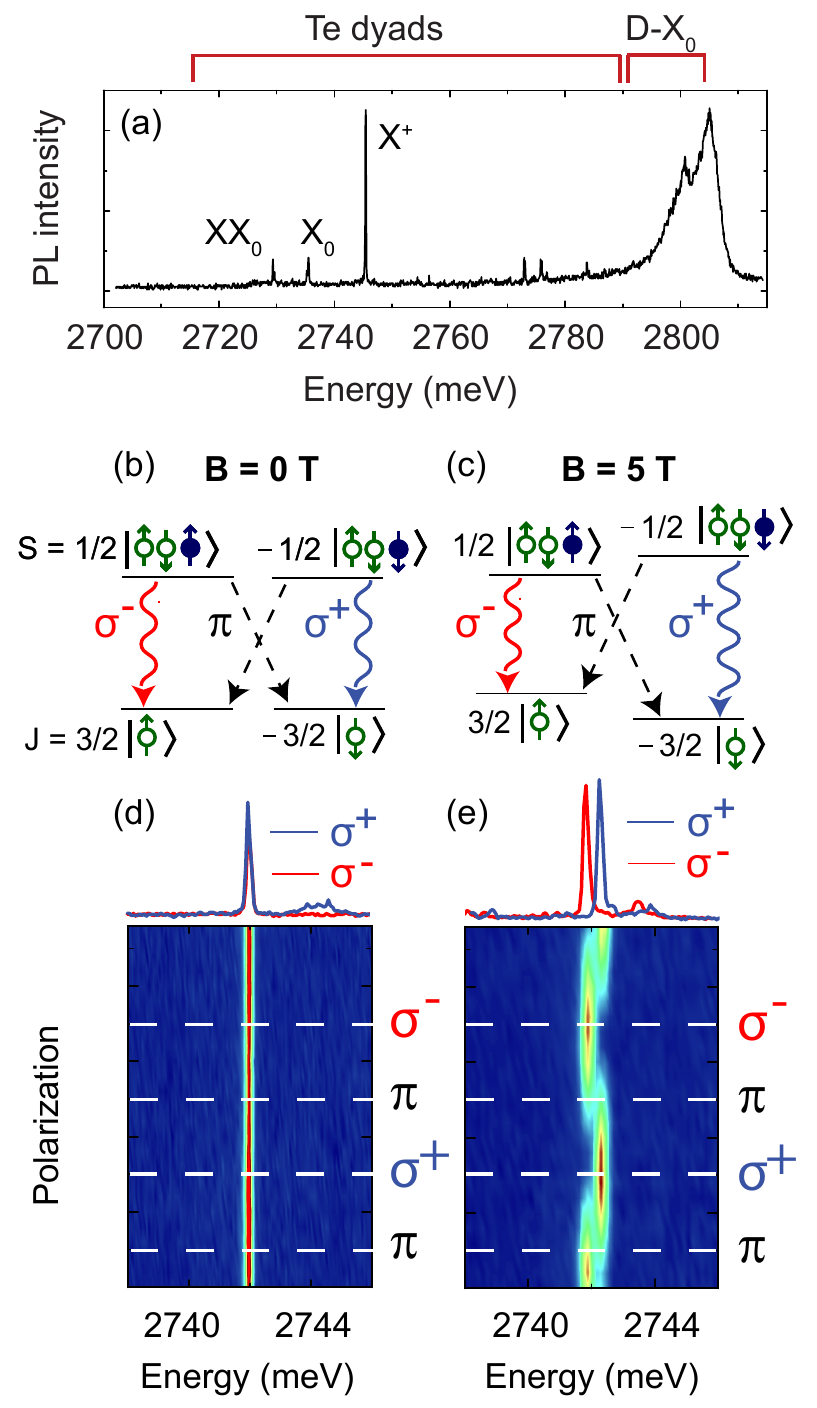}
		\caption{(a) Time-integrated micro-photoluminescence spectrum of the sample studied in this work. This spectrum reveals the presence of a three distinct Te dyads preferentially binding neutral excitons ($X_0$), bi-exciton ($XX_0$), and positive trions ($X^+$). (b) and (c) Schematic diagrams of the selection rules associated to trion emission at $B=0$ and $5~\mathrm{T}$ in a Faraday configuration. $\sigma$ ($\pi$) represent circularly (linearly) polarized transitions, and solid (empty) circles represent electrons (holes). (d) and (e) Trion photoluminescence intensity as a function of energy and polarization of the emission at $\mathrm{B=0}$ and $\mathrm{5~T}$, respectively.}
		\label{zeemanSplitting}
	\end{figure}

	
	Fig. \ref{zeemanSplitting} (a) presents a CW micro-photoluminescence spectrum measured at 4K. The 5~meV wide emission observed slightly above 2800 meV is associated to the radiative recombination of neutral donor bound excitons ($D$-$X_0$)\cite{Merz1972}. The chemical identity of these donors have not been conclusively identified, but may result from aluminium atoms on zinc sites. As demonstrated in Ref. \onlinecite{Kuskovsky2001, Muller2006, Marcet2010}, the emission from Te dyads is found between 2710 to 2790 meV. Emission over this wide energy range is attributed to the existence of several Te dyad configurations with different interatomic separations. This is similar to the well-known emission observed from N dyads in GaP\cite{Thomas1966} or GaAs \cite{Liu1990, Francoeur2004}, where the exciton binding energy varies over a range of about \SI{150}{\milli\electronvolt} with N separation. As this interatomic separation is rigidly set by the anionic sublattice, the emission energy varies in large discrete steps. The inhomogeneous broadening associated with a given dyad configuration is then only limited by its environment. Polarization studies of the exciton emission revealed that most Te dyads in this $\delta$-doped layer have a $C_{2v}$ symmetry with impurities aligned along $<110>$\cite{Muller2006, Marcet2010}. Due to the compressive strain created by the GaAs substrate on the ZnSe layer, light-holes (LH) are pushed \SI{12}{\milli\electronvolt} towards higher energy and only heavy-hole (HH) excitonic complexes are observed. 
	
	Te is a pseudo-donor that can trap an itinerant hole in a short-range potential\cite{Kent2001}, which can then trap through its Coulomb field an electron to form an exciton or an exciton to form a positive trion. The spectrum shown in Fig. \ref{zeemanSplitting}(a) presents three Te dyads binding an exciton ($X_0$), a positive trion ($X^{+}$) and a biexciton ($XX_{0}$). The assignment of the different excitonic complexes was based on the arguments developed in Ref. \onlinecite{Marcet2010a}, related to (1) the dependence between the intensity of the emission and the excitation power, and (2) the fine structure splitting at $B=0~T$, which is expected to vanish for positive trions due to the lack of exchange interaction between the hole singlet and the electron. 
	
	Two key results previously reported suggest that trions are positively charged: their binding energy is positive with respect to the neutral exciton and their diamagnetic shift is only slightly superior to that of the exciton\cite{Marcet2010}. Had the trion been negatively charged, the binding energy would likely have been negative due to the hole-attractive potential of Te and the diamagnetic shift would have been significantly reduced or even negative\cite{Fu2010}, as the final state would be that of a free or very weakly bound electron. As will be discussed later, the polarization memory observed in this work and the high efficiency of the spin initialization process confirms that the extra charge is positive. From the reminder of this work, we describe and analyze the emission from positively charged trions ($X^+$) as the one shown in \ref{zeemanSplitting} (d) and (e).
	
	Optical initialization requires the strong optical selection rules depicted by the strait arrows in Fig. \ref{zeemanSplitting} (b)-(c): the circularly polarized emission, $\sigma^{+}$ or $\sigma^{-}$, of heavy-hole trions ($X^+$) initializes the hole in spin state $J=-3/2$ or $J=+3/2$, respectively. However, this process is efficient only if these hole states are exempt from any admixture from the light-hole states, since mixing add two linearly polarized ($\pi$) emission channels (dashed lines in Panels (b)-(c)) compromising the fidelity of this optical initialization scheme. Two key results indicate that a low valence band mixing despite their energy separation of only 12 meV. First, luminescence under a magnetic field in Faraday configuration does not reveal the presence of additional linear polarized transitions (Fig. \ref{zeemanSplitting} (e)). Second, an analysis of the polarization-resolved PL intensity of exciton states (see Supplementary Information), which present a more complex fine structure than trions due to the non-vanishing exchange interaction, with a model similar to those developed in Ref. \onlinecite{Tonin2012, St-Jean2015} give an LH admixture of 0.6\%. This LH-HH mixing is lower than those typically observed in quantum dots (e.g.  $2 ± 1~\%$ mixing has been measured in self-assembled InAs/GaAs QDs \onlinecite{Tonin2012}), due to lower symmetry-breaking in-plane anisotropic strain fields, and appears to be another distinctive feature of ICs \cite{St-Jean2015, Ethier-Majcher2015}. We next discuss the trion emission with the assumption that the LH-HH mixing has negligible effects. 
	
\begin{figure*}
		\centering
		\includegraphics[trim=0cm 0cm 0cm 0cm, clip]{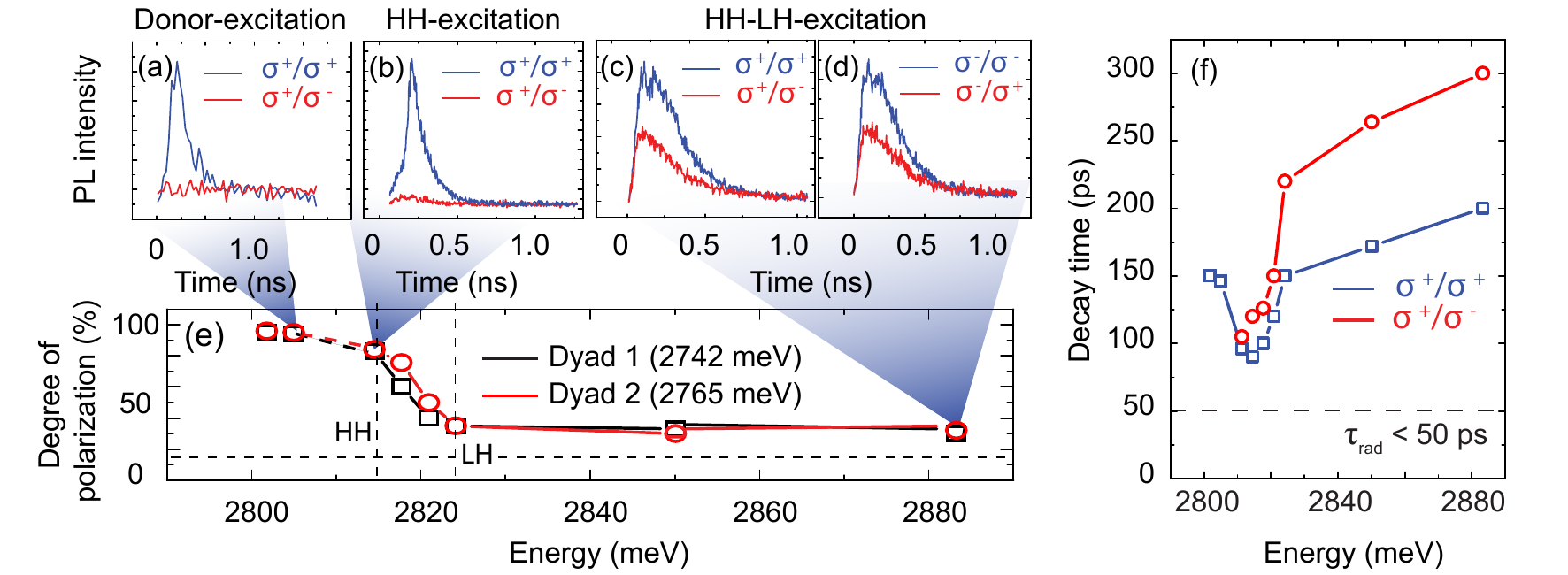}
		\caption{Time-resolved photoluminescence of trions bound to dyad 1 presented in panel (a) of Fig. \ref{polExcitation} for $\sigma^{+}$ excitations tuned (a) to the D-X$^{0}$ band (a) and (b) to the HH band edge, and (c-d) for $\sigma^{+}$ and $\sigma^{-}$ excitations far above the LH band edge but below the spin-orbit band edge. The blue (red) curves show the emission intensity under co-polarized (cross-polarized) circular excitation and detection. (e) The degree of polarization for dyads 1 (black) and 2 (red) is presented as a function of the excitation energy. The horizontal dashed line indicates the degree of polarization of the emission for an excitation energy 250~meV above the ZnSe band gap. (f) Decay times of trion emission for co- (blue squares) and cross-polarized (red circles) configurations as a function of the excitation energy for dyad 1. The dashed line indicates the upper bound for the radiative decay time of trion states ($<50~\mathrm{ps}$)}
		\label{polExcitation}
	\end{figure*}

	
	As shown Fig. \ref{polExcitation}(e), the degree of polarization of the trion emission depends sensitively on the energy of the circularly polarized excitation. Two dyads exhibiting very similar behaviors are presented: Dyad 1 corresponds to the one shown in Fig.\ref{zeemanSplitting}, and Dyad 2 is another dyad of $C_{2v}$ symmetry emitting at a slightly higher energy (2.765 e V). Panels (a) to (d) show the time-dependence of the luminescence of Dyad 1 under co-polarized  ($\sigma^{+}/\sigma^{+}$ or $\sigma^{-}/\sigma^{-}$) and cross-polarized ($\sigma^{+}/\sigma^{-}$ or $\sigma^{-}/\sigma^{+}$) excitation and detection for the three excitation energies indicated in panel (e). The degree of polarization is discussed next, starting from high energy excitation. 
	
	For an excitation energy (\SI{\sim250}{\milli\electronvolt}) above the gap, the trion photoluminescence exhibits a small but non-zero degree of polarization (\SI{15}{\percent}) indicated by the horizontal dashed line. As the excitation energy decreases, the degree of polarization slowly increases and at \SI{2885}{\milli\electronvolt} (\SI{65}{\milli\electronvolt} above the gap), the degree of polarization reaches $30~\%$. Panels (c) and (d) from which the degree of polarization was calculated show that both $\sigma^{+}$ and $\sigma^{-}$ excitations lead as expected to identical degrees of polarization and decay times. At these energies higher than the LH bands but lower than the split-off bands, spin-up and spin-down electrons are generated in a 1:3 ratio (3:1) under a $\sigma^{+}$ ($\sigma^{-}$) excitation\cite{Dyakonov2008}, which limits the degree of polarization to a theoretical maximum of 50~\%. The lower polarization observed results from electron spin randomization in the thermalization, diffusion, and capture processes. 
	
	Decreasing the excitation energy such that only the HH bands are excited significantly increases the degree of polarization: from 30~\% at the LH bands to more than 80~\% at the HH band (Panel (b)). Although exciting below the LH bands is clearly favorable, exciting resonantly at the HH extremum lead to an optimal polarization as momentum-dependent spin-admixture and spin-relaxation processes are minimized. At this band edge, $\sigma^{-}$ ($\sigma^{+}$) photons now only generates spin-up (-down) electrons in the conduction band. The reduced polarization observed implies that some electron spin relaxation processes remain effective. Nonetheless, this very strong polarized emission provides conclusive evidence that trions bound to Te dyads carry a net positive charge. Indeed, the emission polarization from a negative trion (formed from two electrons in a $S=0$ singlet state and a $J=±3/2$ hole) is solely determined by the hole spin state, and should exhibit a  negligible polarization memory due to the very rapid ($T_{1}\sim \mathrm{1~ps}$ \cite{Damen1991}) spin-orbit-induced hole-spin relaxation. In contrast, the emission  polarization from a positive trion is determined by the electron spin state, which is much more robust against spin-orbit-induced relaxation mechanisms, and should therefore exhibit a non-negligible polarization memory.
	
	For an excitation energy below the HH band gap, the optical generation of charge carriers is inhibited by the lack of absorption in the ZnSe layer. However, it is possible to recover efficient trion emission by optically pumping the donor-bound excitons band ($D$-$X_0$) shown in Fig. \ref{zeemanSplitting} (a). At an excitation energy of \SI{2.805}{\electronvolt}, the degree of polarization measured from several dyads is greater than $98.5~\%$, a figure which is limited by the noise associated to the measurement. This high polarization memory results from three processes efficiently preserving the spin of the electron: resonant excitation of donor-bound spin-polarized excitons, exciton tunneling to Te dyads, and trion emission, as discussed below.
	
	Resonant excitation of bound excitons suppresses most spin-orbit related relaxation processes occurring at the HH band edge. Bound electrons have been shown to exhibit long relaxation ($T_{1}\sim 1.6\mu\mathrm{s}$\cite{Heisterkamp2015}) and dephasing ($T_{2}^{*}>30~\mathrm{ns}$\cite{Greilich2012}). For the time during which they are bound to the donor, the exciton spin is preserved because, in contrast to quantum dots, donor exhibit a high symmetry ($T_d$) effectively protecting them from exchange-induced precession mechanisms. The efficient recovery of trion emission indicates that donor-bound excitons rapidly tunnel to nearby Te dyads before their radiative recombination lifetime estimated at $> \SI{200}{\pico\second}$\cite{Greilich2012}. Phonon-assisted tunneling of charges from donors to ICs have been demonstrated in various IC systems\cite{Dean1969, St-Jean2014}, and is expected to be very efficient in ZnSe due to the relatively strong electron-phonon coupling. Our results demonstrate for the first time that this tunneling efficiently preserves the electron spins. Finally, during the short trion lifetime before radiative recombination, the electron spin does not experience significant relaxation and, as discussed above, the HH-LH mixing has a negligible effect of emission selection rules. Following this polarized emission, the hole is bound to the IC in a spin state given by the polarization of the emission as indicated in Fig. \ref{zeemanSplitting} (b). In contrast, to other reported spin initialization schemes, such as coherent population trapping \cite{Xu2008,Brunner2009}, ionization of excitons \cite{Kroutvar2004, Brash2015} and optical pumping \cite{Atature2006, Gerardot2008, Sleiter2013}, our scheme does not require resonant excitations of trion states, special sample structure, or external magnetic fields, but nonetheless achieves near-unity fidelity.
	
	
	Through the decay time of the trion photoluminescence, we demonstrate that this high-fidelity spin-initialization can be achieved on a picosecond timescale. Figure \ref{polExcitation} (f) indicates the decay time of trion emission as a function of excitation energy and polarization configuration. Decay times are obtained from mono-exponential fits of the experimental decay curves such as those presented in Fig. \ref{polExcitation} (a)-(d). For all excitation energies, the decay times are one order of magnitude lower than that of excitons\cite{St-Jean2015, Muller2006}, which is explained by the absence of dark states influencing the dynamics. Nonetheless, these decay times are significantly longer than the spontaneous emission time estimated from the spectral linewidth or Rabi oscillations measurements (not shown, but qualitatively similar to those reported in Ref. \onlinecite{Ethier-Majcher2014} on $\mathrm{GaAs:N_{2}}$), which both indicated a spontaneous lifetime of less than \SI{50}{\pico\second} (indicated with the dashed line in Fig. \ref{polExcitation} (f)). The observed decay time is therefore dominated by processes occurring prior to radiative emission. 
	
	At an excitation energy of \SI{2880}{\milli\electronvolt}, trion decay times are respectively 200 and \SI{300}{\pico\second} for co- and cross-polarized configurations. This significant difference is explained by the coexistence of two factors: 1) different initial electron spin populations (spins are  generated in 3:1 ratio), and 2) the presence of spin-flip mechanisms prior to trion formation on the Te dyad. A balance population model taking into account these two factors reveal a capture time (all processes prior to trion formation) of \SI{180}{\pico\second} and a spin-flip time of \SI{450}{\pico\second}. This dynamic is strikingly different for trions bound to N dyads in GaAs, where the spin-flip rate dominates the capture time, yielding identical decay times irrespective of the polarization configuration. From 2880 to \SI{2825}{\milli\electronvolt}, decay time decreases due to lower excess energy and momentum, and spin-flip time increases as indicated by the higher polarization memory shown in Fig. \ref{polExcitation}.
	
	Below the light-hole band gap, both decay times abruptly shorten. Although there is less energy and momentum to shed, it also becomes increasingly favorable to form excitons instead of electron-hole pairs. Approaching the heavy-hole gap, exciton formation dominates and a capture time of \SI{92}{\pico\second} is recorded. This regime of accelerated capture suggests that the most efficient trion formation mechanism is through whole exciton capture. This mechanism is probably even faster than the value quoted due to the limited time-resolution of the detection system ($\mathrm{80~ps}$), demonstrating that both the capture time and the spontaneous emission lifetime are indeed quite fast. At the heavy-hole gap edge, decay times for both polarization configurations are similar, since the time associated to spin-flip events is now much longer than the capture time. Finally, the decay time of the trion luminescence for resonant excitation of the donor bound exciton ($D$-$X^0$) is \SI{150}{\pico\second}. It is principally determined by exciton tunneling time from the neutral donor to the Te dyad, which is similar for every dyads measured due to the relatively high donor concentration. In contrast to heavy-hole band edge excitation, exciton capture is slower, but it better protects the electron-spin as demonstrated earlier.
	
	These initialization times are several orders of magnitude faster than optical pumping schemes with an external magnetic field in Faraday configuration \cite{Atature2006} ($\mu$s-timescale and fidelities of $99.5~\%$), and one order of magnitude faster than optical pumping schemes in Voigt configuration \cite{Emary2007, Xu2007} (ns-timescale and fidelities of $98.9~\%$). This speed-up allows for the initialization on timescales much shorter than the expected coherence time of hole-spins in ZnSe. In fact, this initialization time is comparable to those obtained through field induced exciton dissociation. This last approach however requires an adapted sample structure for electron ionization, a preselection of quantum dots with low fine-structure splitting, and an elaborate strategy to suppress  exchange-induced spin precessions prior to ionization. \cite{Brash2015}
	
	In summary, we have demonstrated picosecond optical initialization of a hole-spin bound to a Te IC in ZnSe with sufficiently high fidelity ($F>98.5~\%$) and low operation time ($T<150~\mathrm{ps}$) for implementing error correction protocols\cite{Preskill1998}. The efficiency of the initialization scheme used, based on the rapid spin-preserving tunneling of excitons from donor-bound states to Te dyads occupied by a single heavy-hole, arise from the long relaxation time of donor-bound excitons, the very low admixture of light-hole states, and the very short radiative lifetime of IC-bound trions.

	\section{Supplemental Material}
	
	\subsection{Magnitude of light- and heavy-hole mixing}
	
	The fine-structure of the exciton emission allows quantifying the admixture of light-hole states into low-energy states of predominantly heavy-hole character \cite{St-Jean2015, Tonin2012}. For dyads of $C_{2v}$ symmetry with the $C_2$ axis perpendicular to the $\delta$-doped layer (in-plane dyads), the degeneracy of heavy-hole exciton states is lifted, giving rise to two linearly polarized bright states: one polarized along the dyad ($\ket{X}$) and the other polarized perpendicularly to both the dyad and the $C_{2}$ axis ($\ket{Y}$). The exciton wavefunctions can be defined as,
	
	\begin{equation}
		\begin{split}
			\ket{X}&=\sqrt{\left(1-\eta^{2}\right)}~\psi_{\mathrm{hh}}^{(\mathrm{x})}+\eta~\psi_{\mathrm{lh}}^{(\mathrm{x})}\\
			\ket{Y}&=\sqrt{\left(1-\eta^{2}\right)}~\psi_{\mathrm{hh}}^{(\mathrm{y})}+\eta~\psi_{\mathrm{lh}}^{(\mathrm{y})}
		\end{split}
	\end{equation}
	
	\noindent in the presence of HH- and LH-mixing of amplitude $\eta$.   $\psi_{\mathrm{hh,lh}}^{(\mathrm{x,y})}$ are the symmetry-adapted heavy- (hh) and light- (lh) hole wavefunctions \cite{Francoeur2010},
	
	\begin{equation}
		\begin{split}
			\psi_{\mathrm{hh}}^{\mathrm{(x)}}&=\frac{1}{\sqrt{2}}~\left( \beta_{e}\phi_{1}+\alpha_{e}\phi_{4}\right)\\
			\psi_{\mathrm{hh}}^{\mathrm{(y)}}&=\frac{i}{\sqrt{2}}~\left(-\beta_{e}\phi_{1}+\alpha_{e}\phi_{4}\right)\\
			\psi_{\mathrm{lh}}^{\mathrm{(x)}}&=\frac{1}{\sqrt{2}}~\left( \beta_{e}\phi_{3}+\alpha_{e}\phi_{2}\right)\\
			\psi_{\mathrm{lh}}^{\mathrm{(y)}}&=\frac{i}{\sqrt{2}}~\left(-\beta_{e}\phi_{3}+\alpha_{e}\phi_{2}\right),\\
		\end{split}
	\end{equation}
	
	\noindent where $\alpha_{e}$ ($\beta_{e}$) are the spin-up (-down) electron states and $\phi_{i}$ are the four hole states ($\phi_{1}=\ket{3/2;3/2}$, $\phi_{2}=\ket{3/2;1/2}$, $\phi_{3}=\ket{3/2;-1/2}$, and $\phi_{4}=\ket{3/2;-3/2}$). Superscripts refer to the polarization of the emission originating from these states. The emission intensity as a function of the polarization angle $\theta$ is given by,
	
	\begin{equation}
		I^{(x)}(\theta)=\left[\sqrt{1-\eta^{2}}~\mathrm{cos}\theta
		-\frac{\eta}{\sqrt{3}}~\mathrm{cos}\theta\right]^{2},
		\label{intensityX}
	\end{equation}    
	
	\noindent and
	
	\begin{equation}
		I^{(y)}(\theta)=\left[\sqrt{1-\eta^{2}}~\mathrm{sin}\theta
		+\frac{\eta}{\sqrt{3}}~\mathrm{sin}\theta\right]^{2},
		\label{intensityY}
	\end{equation}

	\noindent such that the polarization dependence of the emission can be used to estimate $\eta$
	
	\begin{figure}[h]
		\includegraphics[trim=0cm 0cm 0cm 0cm, clip, width=83mm]{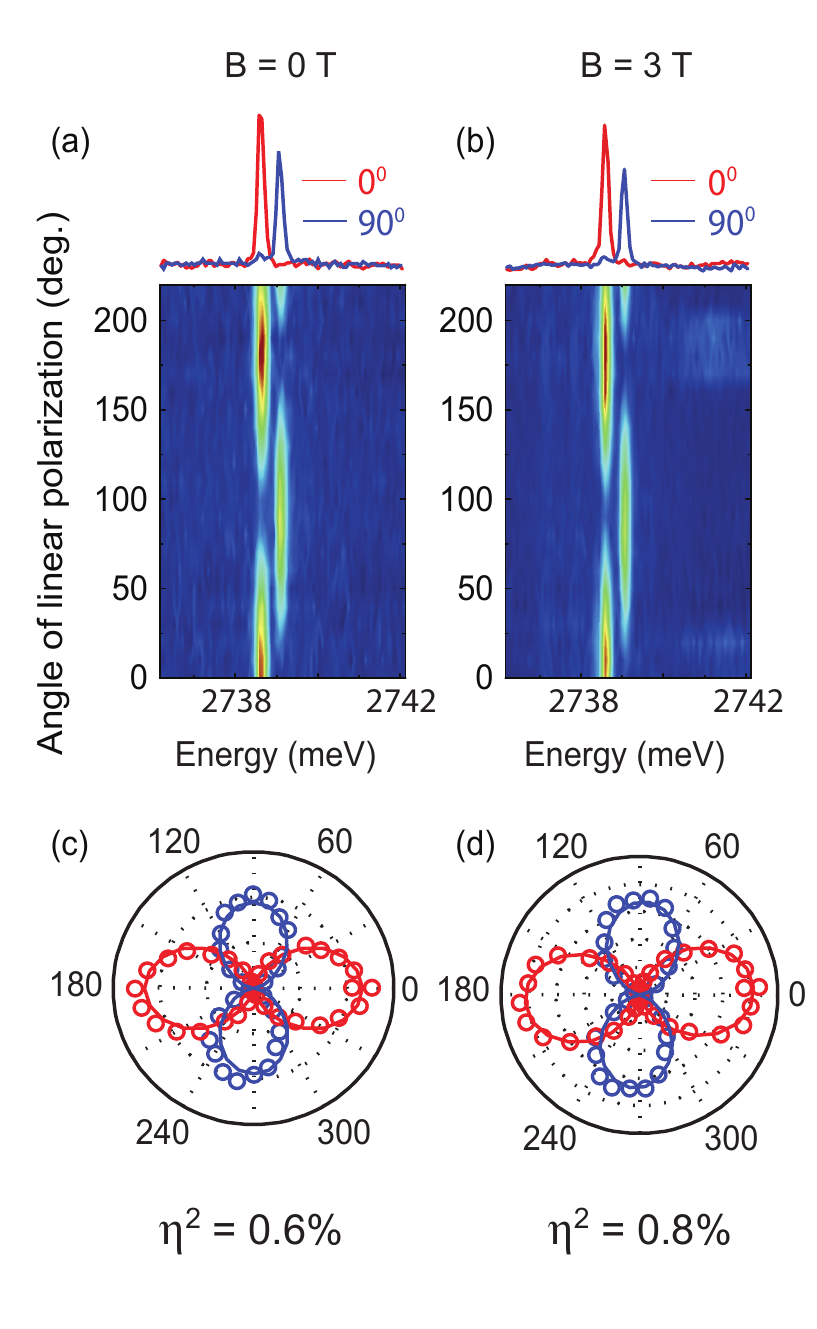}
		\caption{(a) and (b) Photoluminescence spectra and linear polarization maps from a neutral exciton bound to an in-plane dyad of $C_{2v}$ symmetry for magnetic fields of $B$=0 and $B$= 3T applied in Faraday configuration. The red (blue) curves show the photoluminescence intensity polarized parallel (perpendicular) to the dyad. (c) and (d) are polar plots of the intensity of the emission as a function of $\theta$ at $B=0$ and B=$3~\mathrm{T}$.}
		\label{excitonSymmetry}
	\end{figure}
	
	Panels (a) and (b) of Fig. \ref{excitonSymmetry} present the exciton photoluminescence intensities for a magnetic field of $B=0$ (a) and $B=3~T$ (b) applied in the Faraday configuration. Panels (c) and (d) present polar plots of the intensity of the same two transitions as a function of the angle of linear polarization ($\theta$). The full lines show calculated intensities, using Eqs. \ref{intensityX} and \ref{intensityY}, that best fitted the data. The extracted valence band mixing corresponds to $\eta^{2}=0.6~\%$, which is only limited by the uncertainty margin of the fits. Similar or lower values of LH-HH mixing have been obtained for all other in-plane dyads studied. Under a magnetic field in Faraday configuration, no further HH-LH mixing is observed, as expected for an in-plane dyad of $C_{2v}$ symmetry in this magnetic field configuration.

	\subsection{Balance of populations model}
	
	The temporal evolution of the emission from both trion states is obtained by solving differential equations describing the evolution of both the trion populations ($n_{\ket{\Uparrow\Downarrow\uparrow}}$ and $n_{\ket{\Uparrow\Downarrow\downarrow}}$ corresponding to the spin-up and -down trion state), and of the spin populations for the photo-generated free electrons ($n_{\uparrow}$ and $n_{\downarrow}$, corresponding to spin-up and -down electrons) before their capture. It is not necessary to calculate the evolution of the free holes spin populations, because their spin relaxes on time scales orders of magnitude faster than their capture, and therefore they do not influence the polarization of the trion emission.
	
	The equations governing the evolution of the trion populations are the following:
	
	\begin{align}
		\frac{\mathrm{d}n_{\ket{\Uparrow \Downarrow \uparrow}}}{\mathrm{d}t} &= \Gamma_{\mathrm{cap}} n_{\ket{\uparrow}} - \Gamma_{\mathrm{rad}} n_{\ket{\Uparrow \Downarrow \uparrow}}, \nonumber \\
		\frac{\mathrm{d}n_{\ket{\Uparrow \Downarrow \downarrow}}}{\mathrm{d}t} &= \Gamma_{\mathrm{cap}} n_{\ket{\downarrow}} - \Gamma_{\mathrm{rad}} n_{\ket{\Uparrow \Downarrow \downarrow}}, 
	\end{align}
	
	\noindent where $\Gamma_{\mathrm{cap}}$ and $\Gamma_{\mathrm{rad}}$ correspond to the rate of capture and spontaneous emission, and are identical for both trion states. The equations governing the evolution of the spin populations of free electrons ($n_{\uparrow}$ and $n_{\downarrow}$) are:
	
	\begin{align}
		\frac{\mathrm{d}n_{\ket{\uparrow}}}{\mathrm{d}t} &= - \Gamma_{\mathrm{cap}} n_{\ket{\uparrow}} - \Gamma_{\mathrm{trans}} \left( n_{\ket{\uparrow}} + n_{\ket{\downarrow}} \right), \nonumber \\
		\frac{\mathrm{d}n_{\ket{\downarrow}}}{\mathrm{d}t} &= - \Gamma_{\mathrm{cap}} n_{\ket{\downarrow}} - \Gamma_{\mathrm{trans}} \left( n_{\ket{\downarrow}} + n_{\ket{\uparrow}} \right),
	\end{align}
	
	\noindent where $\Gamma_{\mathrm{trans}}$ is the rate of electron spin-flip in the bulk due to spin-orbit interactions. 
	
	The initial conditions of trion populations were set to
	
	\begin{equation}
		n_{\ket{\Uparrow\Downarrow\uparrow}} (t=0) = n_{\ket{\Uparrow\Downarrow\downarrow}} (t=0) = 0,
	\end{equation}
	
	\noindent while those for the free electrons spin populations were dictated by the energy and polarization of the excitation. The populations were normalized to unity, such that:
	
	\begin{align}
		n_{\ket{\downarrow}} (t=0) &= a\\
		n_{\ket{\uparrow}} (t=0) &= 1-a, \nonumber
	\end{align}
	
	\noindent where $a=0.5$ for a linear excitation, $a=1$ ($a=0$) for a $\sigma^{+}$ ($\sigma^{-}$) excitation resonant with the HH valence band, and $a=0.75$ ($a=0.25$) for a $\sigma^{+}$ ($\sigma^{-}$) excitation above both the LH and HH valence bands.
	
	The instantaneous PL intensity of both trion states is given by:
	
	\begin{equation}
		I_{i}(t) = \Gamma_{\mathrm{rad}} n_{i} (t),
	\end{equation}
	
	\noindent and $\Gamma_{\mathrm{rad}}$, $\Gamma_{\mathrm{cap}}$, and $\Gamma_{\mathrm{trans}}$ were adjusted until the calculated PL decay curves ($I_{i}(t)$) best fitted the experimental curves.
	
	
	The values extracted for the different parameters are presented in Table \ref{parameters}, for an excitation tuned with the donor band (D-$\mathrm{X_{0}}$; Fig. 2 (a) of the manuscript) and the HH valence band (HH; Fig. 2 (b)), and far above both the HH and LH valence band (HH-LH; Fig. 2 (c-d)).
	
	\begin{table}[h]
		\centering
		\caption{Extracted parameters for the rates of trion radiative decay ($\Gamma_{\mathrm{rad}}$), capture ($\Gamma_{\mathrm{cap}}$) and spin-flip in the bulk ZnSe ($\Gamma_{\mathrm{trans}}$), for an excitation tuned with the donor -band (D-$\mathrm{X_{0}})$), the HH valence band, and far above both the HH and LH valence bands.}
		\begin{tabular}{ L{2cm} C{1.5cm} C{1.5cm} C{1.5cm} }
			\hline\hline
			&&& \\ [-5pt]
			& D-$\mathrm{X_{0}}$ & HH & HH-LH \\
			\hline
			&\multicolumn{3}{c}{} \\ [-5pt]
			$\Gamma_{\mathrm{rad}}^{-1}$~(ps) & $\mathrm{50\pm10}$ & $\mathrm{50\pm10}$ & $\mathrm{50\pm10}$\\[3pt]
			
			$\Gamma_{\mathrm{cap}}^{-1}$~(ps) & $\mathrm{110\pm10}$ & $\mathrm{90\pm10}$ & $\mathrm{200\pm20}$ \\[3pt]
			
			$\Gamma_{\mathrm{trans}}^{-1}$~(ps) & $\mathrm{0}$ & $\mathrm{670\pm20}$ & $\mathrm{510\pm20}$ \\[2pt]
			\hline\hline
		\end{tabular}
		\label{parameters}
	\end{table}

\end{document}